\documentclass[aps, prb, twocolumn, superscriptaddress, showpacs]{revtex4-1}
\usepackage{graphicx}
\usepackage{amsmath,amssymb}
\begin{document}

\title{Optimizing Hartree-Fock orbitals by the density matrix renormalization group}

\author{H.-G. Luo}
\affiliation{Center for Interdisciplinary Studies and Key Laboratory for Magnetism and Magnetic Materials of MOE, Lanzhou University, Lanzhou 730000, China\\}
\affiliation{Institute of Theoretical Physics, Chinese Academy of Sciences, Beijing 100190, China}
\author{M.-P. Qin}
\affiliation{Institute of Physics, Chinese Academy of Sciences, Beijing 100190, China\\}
\author{T. Xiang}
\email{txiang@aphy.iphy.ac.cn}
\affiliation{Institute of Physics, Chinese Academy of Sciences, Beijing 100190, China\\}
\affiliation{Institute of Theoretical Physics, Chinese Academy of Sciences, Beijing 100190, China}

\begin{abstract}
We have proposed a density matrix renormalization group (DMRG) scheme to optimize the one-electron basis states of molecules. It improves significantly the accuracy and efficiency of the DMRG in the study of quantum chemistry or other many-fermion system with nonlocal interactions. For a water molecule, we find that the ground state energy obtained by the DMRG with only 61 optimized orbitals already reaches the accuracy of best quantum Monte Carlo calculation with 92 orbitals.
\end{abstract}

\maketitle

\section{Introduction} \label{sec1}

Theoretical investigation of correlation effects beyond the Hartree-Fock approximation has long been a challenging problem in the study of
quantum many-body physics. It encounters a number of intractable problems even in the calculation of a helium atom. \cite{Tew2007} The difficulty arises mainly from two respects. One is the approximation, for example the Hartree-Fock approximation, that is used in selecting the one-electron basis states (i.e. molecular orbitals) from a truly infinite basis set. The other is the approximation that is used in further determining the many-electron wavefunction. The full configuration interaction can treat the many-body correlation rigorously. However, the number of orbitals that can be handled by full configuration interaction is small. \cite{Rossi1999} In practical calculation, certain approximations, for example the truncated configuration interaction or the coupled cluster expansion method, \cite{Cizek1980, Bartlett2007} have to be taken.

Recently, the application of the DMRG (Ref.[\onlinecite{White1992}]) has attracted great interest in the quantum chemistry calculation. \cite{Xiang1996, White1999, Dual2000, Chan2002, Chan2003, Legeza2003, Rissler2005} The DMRG is an accurate method for investigating quantum many-body systems. It is variational and has been applied extensively and successfully to the study of strongly correlated electronic materials. In 1999, White and Martin \cite{White1999} made the first DMRG calculation of the ground state energy of water molecule. Within a basis set of 25 Hartree-Fock orbitals, they found that the ground state energy by the DMRG already converges to the exact result obtained from the full configuration interaction \cite{Bauschlicher1986} by just keeping 400 many-body basis states. It reveals the potential of the DMRG in the quantum chemistry calculation. In 2003, Chan and Gordon made a benchmark calculation for the ground state energy of water molecule by using 41 Hartree-Fock orbitals and up to $6000$ many-body bases. \cite{Chan2003} Their result, $-85.512 E_h$ ($E_h$ is the Hartree unit of energy, the attractive energy from nucleus $9.197 E_h$ is included), is by far the most accurate ground state energy of water molecule obtained with 41 Hartree-fock orbitals. A comparable result was also obtained by Legeza and Solyom. \cite{Legeza2003} But it is still much higher than the experimental value.

To improve the accuracy, one can increase both the size of the one-electron basis set and the number of states retained in the DMRG calculation. But this demands a dramatic increase of computer resource. By keeping 6000 states, 41 orbitals are almost the upper limit of one-electron basis set that can be handled by the DMRG with the currently available computers. In this paper, we will show that the accuracy can in fact be more efficiently improved by optimizing the Hartree-Fock molecular orbitals using the DMRG. The idea is that in a real molecular system, the dimension of the Hilbert space is in fact infinite, and the Hartree-Fock orbitals are only a few these basis states selected by the self-consistent Hartree-Fock approximation. This is a single-particle approximation. It underestimates the correlation between electrons. Other orbitals not including in this Hartree-Fock basis set may also have significant contribution to the ground state. To include as much as possible these contributions, the one-electron orbitals need to be re-orthogonalized using a many-body method in a larger Hartree-Fock basis space. This re-orthogonalization optimizes the one-electron basis set and can be carried out using the DMRG. Using 41 orbitals optimized from 92 Hartree-Fock orbitals, for example, we can get a significantly better result for the ground state energy, $-85.558 E_h$, by just keeping up to 500 states in the DMRG calculation. This scheme of optimization can be naturally integrated in the standard DMRG calculation of many-electron systems and allows a large basis set to be optimized. This is different from the canonical transformation, \cite{White2002} the complete active space self-consistent field \cite{Ghosh2008} and other orbital optimization schemes.

The paper is organized as follows. In Section \ref{sec2} we present an explicit scheme for the optimizing the single-particle Hartree-Fock orbitals. In Section \ref{sec3} we take the water molecule as an example to test the efficiency of the scheme proposed and analyze its advantage. In Section \ref{sec4} we compare the ground state energy of the water molecule obtained by the present scheme with the previous results in the literature. Finally, Section \ref{sec5} is devoted to a brief summary and outlook.

\section{Optimization scheme} \label{sec2}

To do the optimization, one needs first to generate a relatively large Hartree-Fock basis set by solving the self-consistent Hartree-Fock equations. The DMRG calculation, however, is done in a subspace of this basis space. In particular, all orbitals will be partitioned into two sets. The first contains all active orbitals that will be used in the DMRG calculation. The second set, which will be taken as a basis reservoir, contains all other orbitals. By solving the Hamiltonian with the DMRG in the active orbital space, one can find the one-electron density matrix from the ground state wave-function. A set of re-orthogonalized orbitals can then be found by diagonalizing the one-electron density matrix. From their occupation numbers, one can identify the contribution of each orbital to the ground state. Both the highest and least occupied orbitals contribute less to the correlation effect. One can freeze these less important orbitals by swapping them with the orbitals in the reservoir. This defines a new set of active orbitals. Again these orbitals can be re-orthogonalized by performing the DMRG calculation. By repeating this procedure many times until and after all the orbitals in the reservoir are activated, the orbitals in the active space will finally become optimized.

Below we take a water molecule to show how this method works. We start by performing a self-consistent Hartree-Fock calculation to find the Hartree-Fock orbitals and the corresponding one- and two-electron integrals. \cite{psi3} The experimental values for the bonding angle between two hydrogens, 104.5 degrees, and the distance between hydrogen and oxygen, 0.957 Angstroms, are used in the calculation. The Hamiltonian for describing a water molecule can then be expressed as
\begin{equation}
H = \sum_{ij,\sigma}t_{ij}c^\dagger_{i\sigma} c_{j\sigma} + \sum_{ijkl,\sigma\sigma'} V_{ijkl} c^\dagger_{i\sigma} c^\dagger_{j\sigma'}c_{l\sigma'} c_{k\sigma}
\label{Ham1}
\end{equation}
where $c^\dagger_{i\sigma}$ ($c_{i\sigma}$) is the creation (annihilation) operator of electron at the $i$th orbital with spin $\sigma$. $t_{ij}$ is the one-electron integral. $V_{ijkl}$ is the tensor for describing the Coulomb interaction between electrons.

Now we divide the Hartree-Fock basis set into the active orbitals in the space $\cal{A}$ and the remaining as reservoir denoting as the space $\cal{I}$. The orbitals in the reservoir are either fully occupied or not occupied by electrons if their energies are below or above the Fermi level. By freezing the orbitals in the reservoir, one can rewrite the Hamiltonian as
\begin{eqnarray}
&&H = E_0  + \sum_{ij\in {\cal A},\sigma}t^{\text{a}}_{ij, \sigma} c^\dagger_{i\sigma} c_{j\sigma} \nonumber \\
&& \hspace{1cm} + \sum_{ijkl\in {\cal A},\sigma\sigma'} V_{ijkl}c^\dagger_{i\sigma} c^\dagger_{j\sigma'}c_{l\sigma^\prime} c_{k\sigma},
\label{Ham2}
\end{eqnarray}
where $E_0$ is a constant energy contributed from all orbitals in the reservoir
\begin{eqnarray}
&& E_0 = \sum_{ij\in \cal{I},\sigma} \left( t_{ii}\delta_{ij} - V_{ijji}  \langle c^\dagger_{j\sigma} c_{j\sigma}  \rangle \right) \langle  c^\dagger_{i\sigma} c_{i\sigma} \rangle
\nonumber \\
&& \hspace{1cm}+ \sum_{ij\in \cal{I}, \sigma\sigma'} V_{ijij}  \langle c^\dagger_{j\sigma'} c_{j\sigma'} \rangle  \langle  c^\dagger_{i\sigma} c_{i\sigma} \rangle .
\end{eqnarray}
$\langle c^\dagger_{i\sigma} c_{i\sigma} \rangle = 1$ if orbital $(i,\sigma)$ is below the Fermi level or $0$ otherwise. $E_0=0$ if all the orbitals below the Fermi level are included in the active space. $t^a_{ij}$ is the one-electron integral between the active orbitals. It includes the contribution from the potential energy between active orbitals and frozen ones in the reservoir:
\begin{equation}
t^{\text{a}}_{ij, \sigma} = t_{ij} + \sum_{k\in {\cal I}, \sigma'}\left( V_{ikkj} - V_{ikjk} \delta_{\sigma\sigma'} \right)
\langle c^\dagger_{k\sigma^\prime}  c_{k\sigma^\prime} \rangle .
\end{equation}

The second term in Eq.~(\ref{Ham2}) is difficult to treat in the DMRG calculation because it is a sum of $O(N^4)$ operators. It is practically infeasible to calculate and store independently all the matrix elements of these operators. To overcome this difficulty, the regrouping technique of operators proposed by Xiang should be used. \cite{Xiang1996} This can reduce the number of independent operators whose matrix elements need to evaluated from $O(N^4)$ to the order of $O(N^2)$.

In the active space $\cal{A}$, the Hamiltonian can be diagonalized by the DMRG. From the ground state wavefunction obtained, $|\psi\rangle$, the single-particle density matrix
\begin{equation}
\rho_{ij} = \langle \psi|c^\dagger_{i\sigma} c_{j\sigma} |\psi\rangle
\end{equation}
can be evaluated. The eigenvectors of $\rho$ define a new set of orthogonal one-electron basis states, called natural orbitals. The eigenvalue of $\rho$ is the occupation number, which measures the probability of the corresponding eigenvector in the ground state. The orbitals with the highest or lowest occupation number have the least contribution to the exchange and correlation energy. They are less important in comparison with other orbitals. Thus by diagonalizing the single-particle density matrix, one can optimize the basis states and order them according to their contribution to the many-body ground state. This completes the first step of orbital optimization. After that, a few of least important orbitals are swapped with the orbitals in the reservoir. The Hamiltonian for the active orbitals is then updated.

The above procedure of optimization can be repeated until all the orbitals in the reservoir are activated. This completes a full cycle of optimization. Generally a few cycles are needed in order to obtain the most optimized orbitals. Once the optimized orbitals are determined, the DMRG iterations with finite lattice sweeping will be performed to find the ground state energy. Below we consider explicitly the Pople-type bases to test the efficiency of the scheme proposed above.

\section{Test of optimization algorithm} \label{sec3}

\begin{figure}[tbp]
\includegraphics[width = \columnwidth]{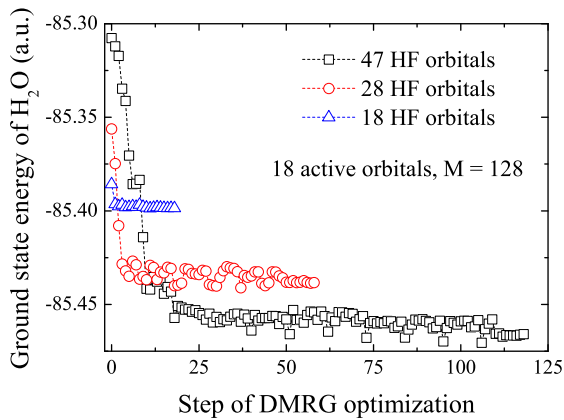}
\caption{(Color online) The ground state energy of H$_2$O as a function of the optimization step for the three sets of Pople-type basis states, which contain $47$, $28$, and $18$ Hartree-Fock orbitals, respectively. There are 18 orbitals in the active space. In the DMRG calculation, $M = 128$ many-body basis states are retained. } \label{fig1}
\end{figure}

Figure \ref{fig1} shows how the ground state energy of water molecule varies with the step of optimization for the 18 active molecular orbitals by the DMRG. Each point in the figure represents a cycle of DMRG calculation within a set of active obtials. Three sets of Pople-type Hartree-Fock bases, 6-311++G(2d, 2p), 6-311+G$^{*}$ and 6-31G$^{*}$, are used. They contain $47$, $28$, and $18$ Hartree-Fock orbitals, respectively. The active space includes all $10$ electrons of H$_2$O. For the basis set 6-31G$^{*}$ with $18$ Hartree-Fock orbitals, all $18$ orbitals are used to perform the DMRG calculation. In this case, the orbital optimization is to recombine the molecular orbitals through the unitary transformation defined by the single-particle density matrix. For the other two cases shown in Fig.~\ref{fig1}, there are orbital exchanges between active and reservoir spaces. At each time $3$ least occupied orbitals in the active space are swapped with the orbitals in the reservoir. A full cycle of optimization needs $4$ and $10$ times of swapping for the systems with 28 and 47 Hartree-Fock orbitals, respectively. Fig. \ref{fig1} shows the results for $10$ full cycles of optimizations of orbitals, and additional $10$ times of the DMRG sweeping in the optimized active space.

For all the three cases shown in Fig.~\ref{fig1}, the orbital optimization improves significantly the DMRG results. The improvement is more striking at the first cycle of orbital optimization. After that, the improvement becomes relatively small. This is because all orbitals have already been activated in the first full cycle of optimization. It suggests that in practical application, two to three cycles of orbital optimization are enough.

\begin{figure}[tbp]
\includegraphics[width = \columnwidth]{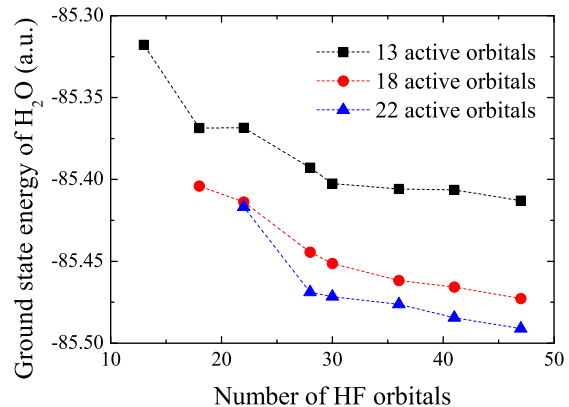}
\caption{(Color online) The ground state energy of H$_2$O as a function of the number of Hartree-Fock orbitals used for optimization. The active space contains 13, 18 and 22 orbitals, respectively. In the orbital optimizations, $M = 128$ states are retained.  In the calculation of the ground state energy from the optimized orbitals using the finite-lattice algorithm of DMRG, the number of states retained is also $M=128$ except in the last step of iteration where $M = 500$ are retained.} \label{fig2}
\end{figure}

Another feature revealed by Fig.~\ref{fig1} is that the more the Hartree-Fock orbitals are used for optimization, the lower (hence better) the ground state energy can be obtained. This is natural since a larger basis set involves more correlation that is underestimated by the Hartree-Fock approximation. To see this more quantitatively, we show in Fig.~\ref{fig2} the ground state energy of H$_2$O as a function of the number of Hartree-Fock orbitals used for optimization. For the three sets of data shown in the figure, which are obtained with 13, 18, and 22 active orbitals respectively, the ground state energy varies almost linearly with the number of Hartree-Fock orbitals. It indicates that the optimization is indeed important. Moreover, the time needed for optimization just scales linearly with the number of Hartree-Fock orbitals, thus the optimization is quite efficient. Allowing the memory space for storing the matrix elements of $V_{ijkl}$, this suggests that as many as Hartree-Fock orbitals should be included in the optimization.

\begin{figure}[h]
\includegraphics[width = \columnwidth]{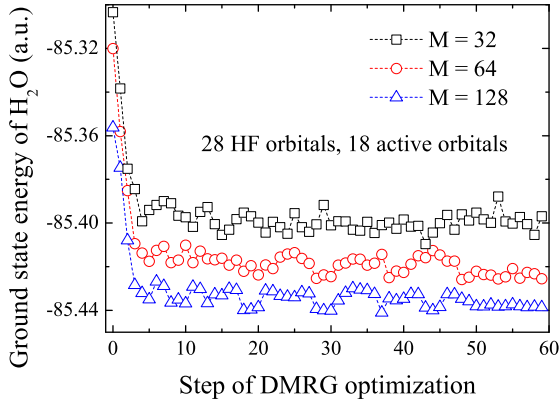}
\caption{(Color online) The ground state energy of H$_2$O obtained by the DMRG with 18 active orbitals, optimized from 6-311+G$^*$ with $28$ Hartree-Fock orbitals. $M$ is the number of states retained in the DMRG calculations.
} \label{fig3}
\end{figure}

The orbital optimization can be also improved by increasing the number of states retained in the DMRG calculation $M$. Fig. \ref{fig3} shows how the ground state energy of H$_2$O varies with the step of optimization by keeping $M = 32$, 64, and 128 states in the DMRG iteration, respectively. The improvement is indeed quite significant when $M$ is increased from 32 to 64, and to 128. But further increasing $M$, more improvement can be achieved. But the speed of improvement will become smaller and smaller, since the ground state energy will converge exponentially with $M$ for sufficiently large $M$. If both computer time and memory space are allowed, the value of $M$ used in the optimization should be taken such that it is just before the ground state energy begins to converge exponentially. Having shown the advantage of the scheme, we apply it to calculate the ground state energy of water molecule and compare it with the results in the literature.

\section{Improved ground state energy of water molecule} \label{sec4}

\begin{figure}[h]
\includegraphics[width = \columnwidth]{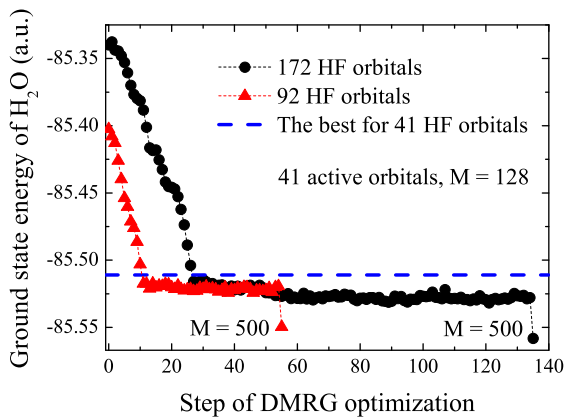}
\caption{(Color online) The ground state energy of H$_2$O obtained with 41 active orbitals, optimized from 92 and 172 Hartree-Fock orbitals, respectively. $M=128$ states are retained in the DMRG calculations, except at the final step at which 500 states are retained. The dash line is the best result of Chan and Gondon, obtained using 41 orbitals and $M=6000$ states.\cite{Chan2003}}\label{fig4}
\end{figure}

The best DMRG result for the ground state energy of water molecule so far available is that obtained by Chan and Gordon in 2003, by using 41 Hartree-Fock orbitals and keeping $M = 6000$ states \cite{Chan2003}. Their value, $-76.314715(E_h)$, sets a variational bound for the ground state energy of H$_2$O. However, by taking $41$ active orbitals optimized from a larger Hartree-Fock basis set, we find that this variational bound can be significantly lowered by keeping a few hundred basis states in the DMRG calculation. Fig.~\ref{fig4} shows the DMRG results for the ground state energy with 41 orbitals, optimized from $92$ and $172$ Hartree-Fock orbitals, respectively. By keeping just $128$ states in the DMRG calculation, we find that the result is already better than that obtained by Chan and Gordon. In the last step of DMRG calculation, the number of states retained is increased from $128$ to $500$. This leads to a sharp drop in the data, further lowering the energy by about 0.05$E_h$. This result shows the potential of orbital optimization in quantum chemistry calculations, since the computer cost in obtaining these results is much smaller than in the calculation of Chan and Gordon. \cite{Chan2003}

To further improve the result, we calculate the ground state energy of H$_2$O by increasing the size of active space to 61 orbitals, optimized from 172 Hartree-Fock orbitals. By keeping 300 states at the last step of DMRG iterations and 150 states in all other steps, we find that ground state energy is $-85.567(E_h)$ which, as shown in Tab.~\ref{table1}, is comparable to the best quantum Monte Carlo result as well as the coupled cluster expansion results obtained from 92 orbitals. But our result is variational.

\begin{table}[h]
\caption{\label{table1} Comparison of the ground state energy of H$_2$O obtained by different methods. Our DMRG result is obtained by using 61 active orbitals optimized from 172 Hartree-Fock orbitals. In all the iteration steps of DMRG, except the last one, $M = 150$ states are retained. In the last step, 300 states are retained. }
\begin{ruledtabular}
\begin{tabular}{lcc}
Method & Number of orbitals & Energy ($E_h$) \\
\hline
HF & 92 & -85.256 \\
CCSD(T) & 92 & -85.563 \\
QMC (Ref.~[\onlinecite{Saidi2006}]) & 92 & -85.567 \\
DMRG (Ref.~[\onlinecite{Chan2003}]) & 41 & -85.512 \\
DMRG (present work) & 61 &  -85.567 \\
\end{tabular}
\end{ruledtabular}
\end{table}

\section{Summary and outlook} \label{sec5}

 In this paper we have shown that the orbital optimization is important in the quantum chemistry calculation. We present a novel DMRG scheme to optimize the Hartree-Fock orbitals. It allows more than 100 orbitals to be treated and improves greatly the accuracy of the results. With $41$ optimized orbitals and $128$ basis states, our DMRG result for the ground state energy of H$_2$O is already better than that reported by Chan and Gordon\cite{Chan2003} with 41 HF orbitals and 6000 states. We find that the ground state energy is -85.567$E_h$ by using 61 optimized orbitals. This result is comparable to the best values reported by the CCSD(T) and quantum Monte Carlo calculations with $92$ Hartree-Fock orbitals. It can be further improved by optimizing orbital orders. \cite{Legeza2003} These optimized orbitals can be used not just by the DMRG, but also by other many-body numerical methods.

\begin{acknowledgments}
We wish to thank C. Guo for the help in the program optimization. Support from CMMM of Lanzhou University, the NSF-China, the program for NCET and the national program for basic research of China is acknowledged.
\end{acknowledgments}

\end{document}